\documentclass[
 reprint,
 groupedaddress,
 twocolumn,
 showpacs,
 amsmath,amssymb,
 aps,
 prl
]{revtex4-1}

\usepackage{graphicx}
\usepackage{dcolumn}
\usepackage{bm}

\usepackage[hidelinks]{hyperref}
\usepackage[all]{hypcap}

\hypersetup{
  colorlinks   = true, 
  urlcolor     = blue, 
  linkcolor    = blue, 
  citecolor    = blue 
}

\begin{document}

\title{Paving the way for third harmonic generation\\in hollow-core photonic bandgap fibers}

\author{Z. Montz$^{1,2}$, and A. A. Ishaaya$^2$}

\affiliation{
$^1$Electro-Optics Unit, Ben-Gurion University of the Negev, Beer Sheva, 84105, Israel\\
$^2$Department of Electrical and Computer Engineering, Ben-Gurion University of the Negev, Beer Sheva, 84105, Israel
}


\begin{abstract}
We present two novel hybrid photonic structures made of silica that possess two well-separated frequency bandgaps. The addition of interstitial air holes in a precise location and size allows these bandgaps to open up with a ratio of ${\sim3}$ between their central frequencies at the air line ${ck_z/w=1}$, thus fulfilling the basic guidance condition for third harmonic generation in hollow-core fibers. In addition, these designs may serve as high-power laser delivery of two well-separated wavelengths, such as visible and near infrared.
\end{abstract}


\maketitle

\thispagestyle{empty}
\pagestyle{empty}


In quantum mechanics, Schr{\"o}dinger's equation predicts that a periodic potential in a conducting crystal will prevent electrons with specific energies from propagating in certain directions. Similarly, in a classical framework, Maxwell's equations predict that a periodic dielectric material will prevent light with specific frequencies from propagating in certain directions. Over the past few decades, periodic dielectric materials have completely revolutionized our control over radiation, leading to the development of a new class of optical fiber light sources termed photonic crystal fibers~(PCFs)~\cite{yeh_theory_1978,knight_all-silica_1996}.

One of the recent breakthroughs in optics was the first demonstration of how light can be guided in a hollow-core PCF (\mbox{HC-PCF}) using a bandgap mechanism~\cite{atkin_full_1995,cregan_single-mode_1999}. This unique \mbox{HC-PCF}, also known as a hollow-core photonic bandgap fiber (\mbox{HC-PBG}), minimizes bulk material losses and nonlinearities that are present in solid-core fibers. Although \mbox{HC-PCFs} typically have a weaker Kerr effect than conventional fibers, introducing gas into the hollow-core and exploiting the long interaction length of the highly confined fundamental mode enhances nonlinear processes.

Recently, it has been demonstrated that Ar-filled kagome \mbox{HC-PCFs} can employ the long interaction length of a Ti:sapphire pump laser source to generate tunable deep-UV ultrafast pulses~\cite{holzer_4_2010,joly_bright_2011,mak_tunable_2013}. Deep-UV laser sources are necessary for various applications that require high spatial coherence. The low-loss at both the pump wavelength and the generated UV wavelength is crucial for efficient conversion to UV.

\begin{figure}[b]
\includegraphics[width=0.41\textwidth]{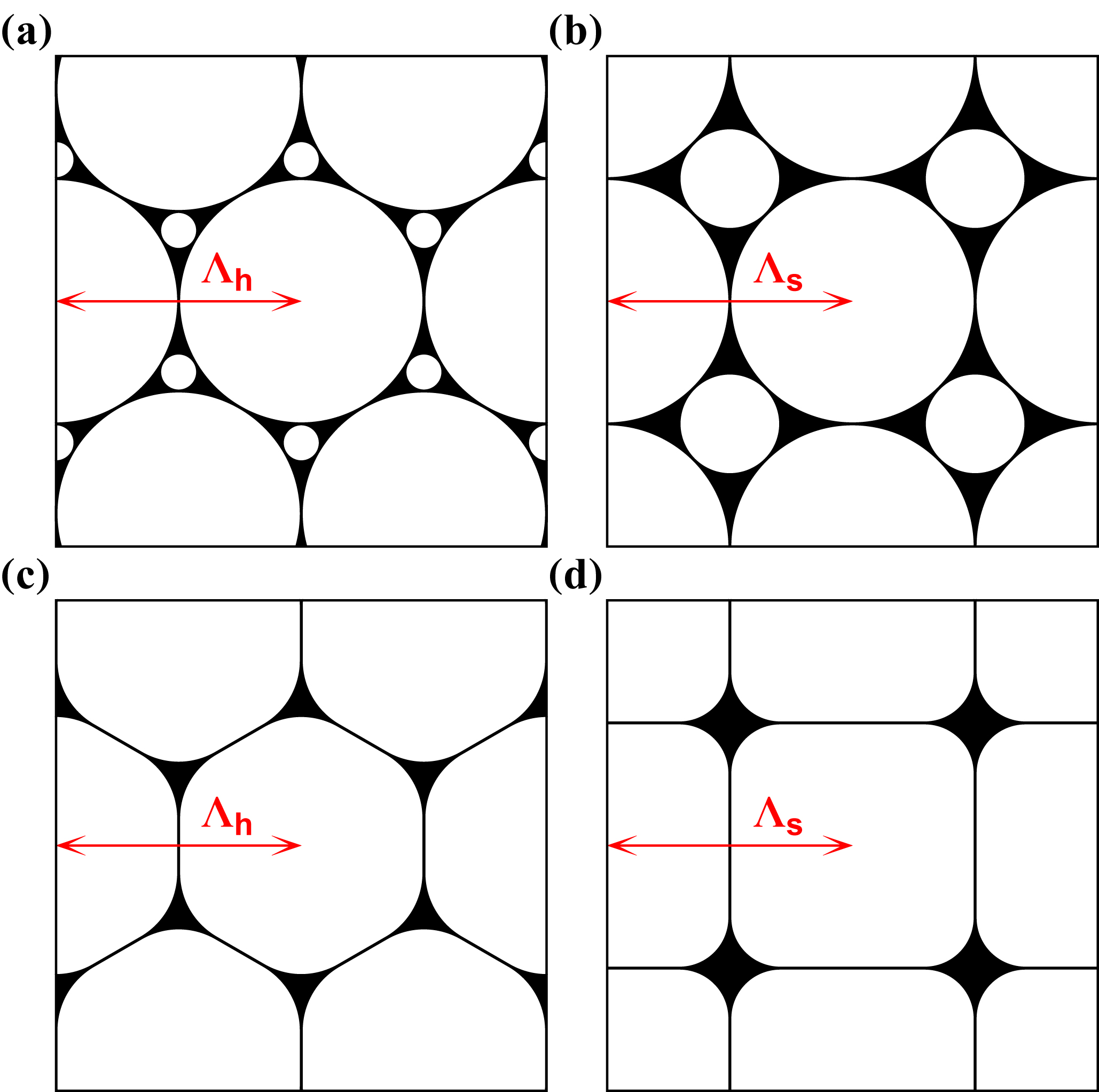}
\caption{\label{fig:1}Cladding structures possessing two frequency bandgaps. Hexagonal (a) and square (b) cladding structures with interstitial air holes at the symmetry points of high-index regions. Hexagonal (c) and square (d) cladding structures with a high-air-filling fraction. ${\Lambda}$ is the lattice constant.}
\end{figure}

Unlike the broadband hollow-core kagome fiber, most \mbox{HC-PBGs} cladding structures thus far reported possess only one restricted bandgap, and are therefore not suitable for guiding light in two well-separated spectral ranges. By adding interstitial holes to fundamental lattices~(Fig.~\hyperref[fig:1]{\ref*{fig:1}(a,b)})~\cite{broeng_highly_1998} or by changing the parameters of high-air-filling structures~(Fig.~\hyperref[fig:1]{\ref*{fig:1}(c,d)})~\cite{mortensen_modeling_2004,poletti_hollow-core_2007,light_double_2009,lyngso_7-cell_2009}, two bandgaps can be created in \mbox{HC-PBGs}, through which radiation can propagate with relatively low losses. A high-air-filling fraction in the silica cladding is typically obtained after pressurizing the stacked capillaries during the drawing process to reduce the final strut wall thickness. In structures previously reported~\cite{broeng_highly_1998,mortensen_modeling_2004,poletti_hollow-core_2007,light_double_2009,lyngso_7-cell_2009}, the ratio between the central frequencies of both bandgaps at the air line is less than 2; therefore, such structures are not capable of guiding both the fundamental and the harmonic waves in a third harmonic generation (THG) process.

Moreover, guiding both the fundamental and the higher harmonic modes in two different bandgaps is not sufficient for THG. Without appropriate phase-matching of these modes, the conversion efficiency is expected to be very low~\cite{agrawal2006nonlinear}.

Several methods have been proposed to phase-match THG in PCFs. For example, a modified total internal reflection (MTIR) mode can be phase-matched with a photonic bandgap mode in a doped solid core PCF~\cite{betourne_design_2008}. However, MTIR modes do not exist in hollow-core fibers, and therefore this method cannot be implemented in \mbox{HC-PBGs}. The kagome fiber, in contrast, presents a potential path towards THG in \mbox{HC-PCFs} because the phase-matching condition can be fulfilled with two different order modes by counterbalancing the kagome dispersion with the gas dispersion~\cite{nold_pressure-controlled_2010}. Previous experiments with a kagome fiber proved beneficial for generating tunable UV ultrafast pulses, but were less promising for THG as the fiber displayed low modal overlap, UV attenuation and group velocity walk-off~\cite{nold_pressure-controlled_2010}. In addition to these drawbacks, exploiting the intensity profile of the third harmonic wave is apparently difficult, as the higher-order mode does not have a Gaussian profile.

Here we propose and investigate two novel cladding designs, demonstrating that \mbox{HC-PBGs} can guide two Gaussian-like modes with high spatial overlap in two well-separated bandgaps suitable for THG. These designs may allow, for the first time, to guide a common laser source (such as Nd:YAG or Ti:sapphire) in \mbox{HC-PBGs} while simultaneously supporting a low-loss spectral window in the UV. We confirm the low-loss regions in both fiber designs numerically with two rigorous full-vectorial frequency domain methods.

\begin{figure}[b]
\includegraphics[width=0.41\textwidth]{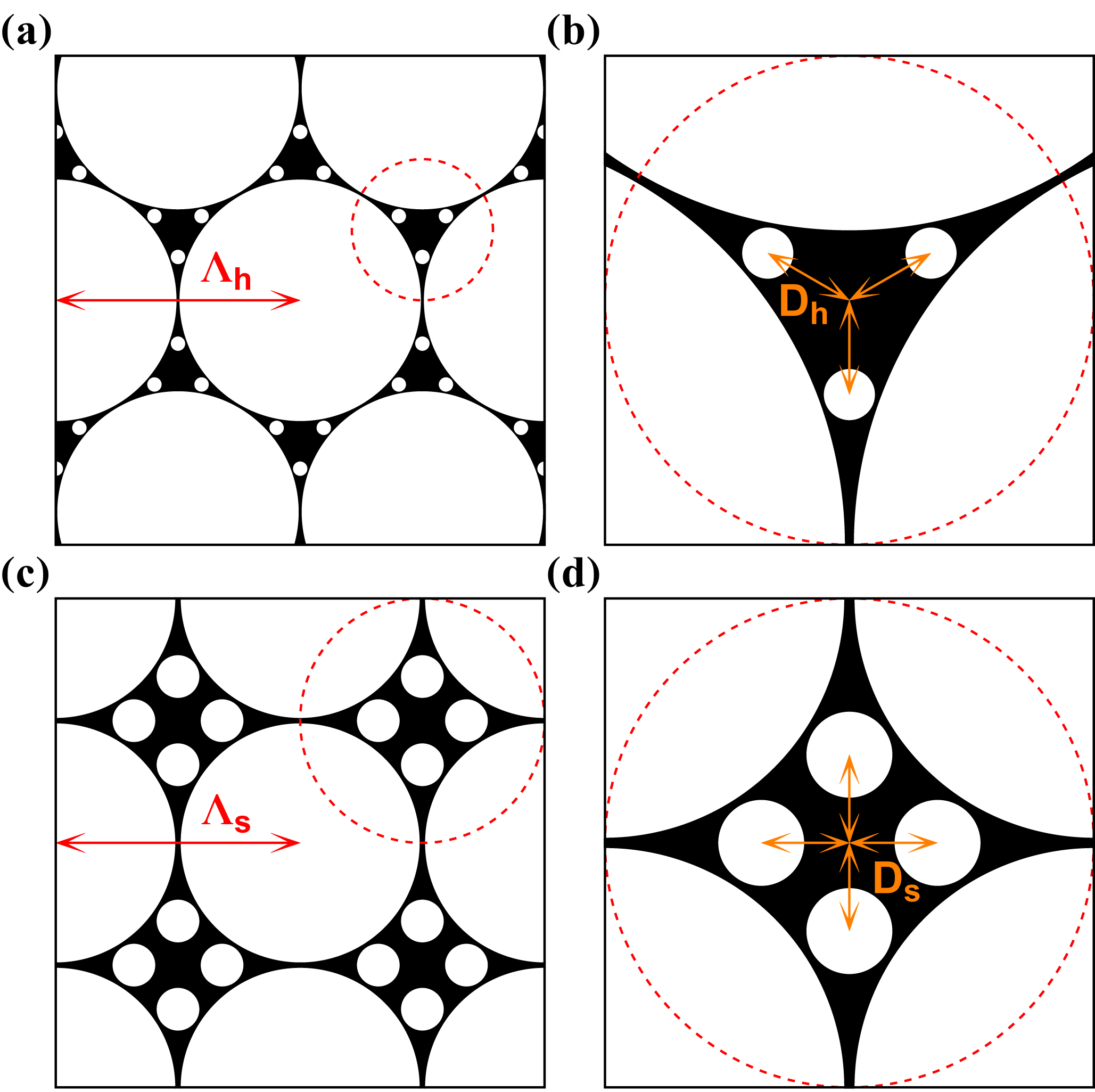}
\caption{\label{fig:2}Hybrid hexagonal (a,b) and square (c,d) cladding structures for THG. ${D}$ is the displacement of the interstitial holes from the symmetry points; ${\Lambda}$ is the lattice constant.}
\end{figure}

When considering an infinite cladding structure, adding interstitial holes to simple hexagonal or square lattices at the common symmetry points of high-index regions~(Fig.~\hyperref[fig:1]{\ref*{fig:1}(a,b)}) fails to create bandgaps in the appropriate location for harmonic generation. Surprisingly, isolating high-index regions of these simple lattices with multiple interstitial holes can open up two bandgaps, wherein the ratio between their central frequencies at the air line is approximately 3.

The two hybrid structures that we hereby propose~(Fig.~\hyperref[fig:2]{\ref*{fig:2}}) are based on hexagonal and square lattices. For the hybrid hexagonal lattice~(Fig.~\hyperref[fig:2]{\ref*{fig:2}(a,b)}), the radius of the large and small air holes are ${r_h^1=0.495\Lambda_h}$ and ${r_h^2\sim0.0305\Lambda_h}$, respectively, where ${\Lambda_h}$ is the hexagonal large air holes pitch. The displacement of the small interstitial air holes from the symmetry points is ${D_h\sim0.1113\Lambda_h}$. Similarly, the geometric parameters of the hybrid square lattice~(Fig.~\hyperref[fig:2]{\ref*{fig:2}(c,d)}) are ${r_s^1=0.49\Lambda_s}$, ${r_s^2\sim0.0884\Lambda_s}$ and ${D_s\sim0.1804\Lambda_s}$, where ${\Lambda_s}$ is the square large air holes pitch.

\begin{figure}[b]
\includegraphics[width=0.41\textwidth]{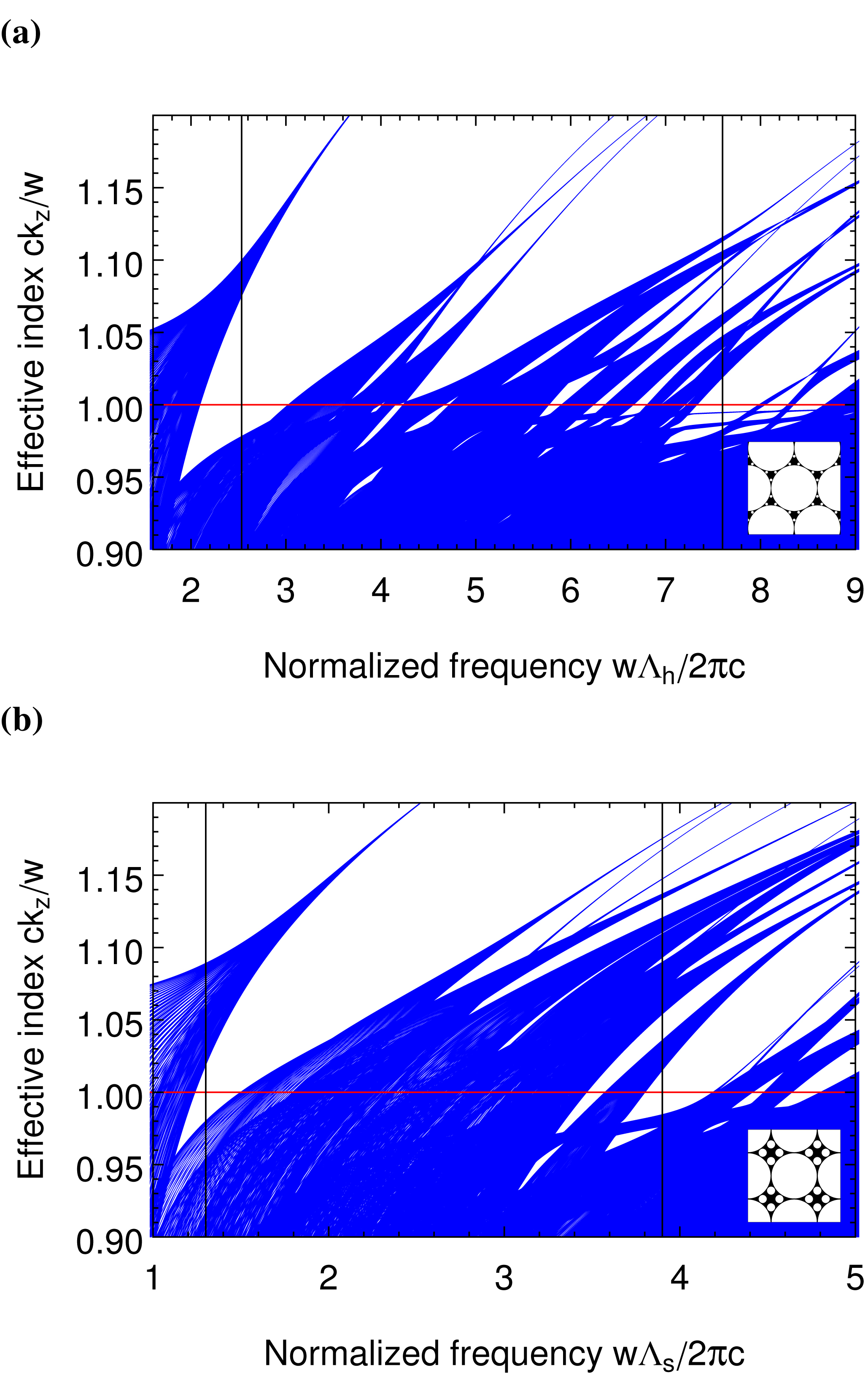}
\caption{\label{fig:3}Band diagrams of the hybrid hexagonal (a) and square (b) structures. In each plot, the red horizontal line corresponds to the air line and the vertical black lines correspond to an optional configuration for guiding a fundamental source and its third harmonic. The vertical lines correspond to the normalized frequencies 7.6 and 7.6/3 in (a), and to 3.9 and 3.9/3 in (b).}
\end{figure}

Band diagrams corresponding to the structures in Fig.~\hyperref[fig:2]{\ref*{fig:2}} are shown in Fig.~\hyperref[fig:3]{\ref*{fig:3}}. Both band diagrams were computed with the plane wave expansion method~(PWE)~\cite{johnson_block-iterative_2001,guo_simple_2003}, with a background dielectric constant set to 2.1. The bandgaps at the higher normalized frequency are more sensitive to variations in the periodic structure than the fundamental gaps; we therefore focused on maximizing the gap-midgap ratio~\cite{joannopoulos2008photonic} of the higher-order bandgaps.

\begin{figure}[b]
\includegraphics[width=0.41\textwidth]{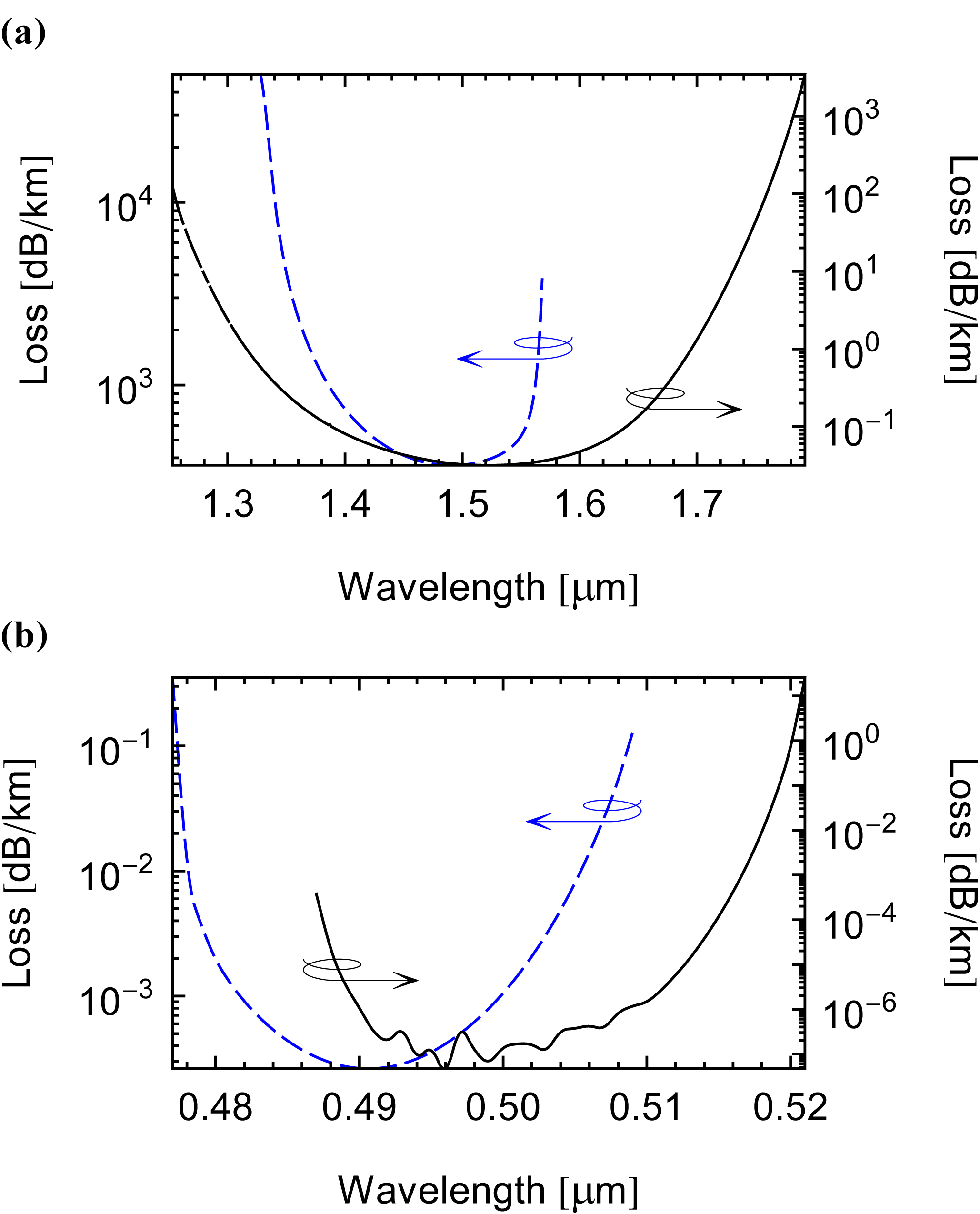}
\caption{\label{fig:4}Confinement loss of the hybrid hexagonal (black) and square (blue dashed) fibers in the fundamental (a) and higher order (b) bandgaps.}
\end{figure}

Figure~\hyperref[fig:3]{\ref*{fig:3}} demonstrates that the first bandgap in the hybrid hexagonal structure opens at a higher normalized frequency than that of the hybrid square structure, resulting in a longer period. The smallest feature size in both structures is the strut thickness. Therefore, the shorter lattice constant in the hybrid square layout leads us to set strut-period ratios of the square structure ${t_s^1/\Lambda_s}$ and ${t_s^2/\Lambda_s}$ to a larger value than the ratios for the hexagonal structure ${t_h^1/\Lambda_h}$ and ${t_h^2/\Lambda_h}$. In both layouts, ${t^1}$ is the strut thickness between neighboring large air holes, and ${t^2}$ is the strut thickness between neighboring large and small air holes.

One may assume that the scalability of periodic dielectric materials allows these structures to support THG with an arbitrary fundamental source by merely changing the lattice constant~${\Lambda}$. However, this assumption is far from accurate because the scalability of the structures shown in Fig.~\hyperref[fig:2]{\ref*{fig:2}} is only valid when the permittivity is a function of the spatial variables~\cite{joannopoulos2008photonic}. Although the permittivity is not frequency-independent, the band diagrams in \mbox{HC-PBGs} are usually less sensitive to the material dispersion and are largely dominated by the geometric waveguide dispersion. For this reason, we first computed the band diagrams with a constant permittivity in an infinite, scalable system. We then used the finite element method (FEM) to investigate the finite features of both hybrid fibers with material dispersion.

\begin{figure}[b]
\includegraphics[width=0.41\textwidth]{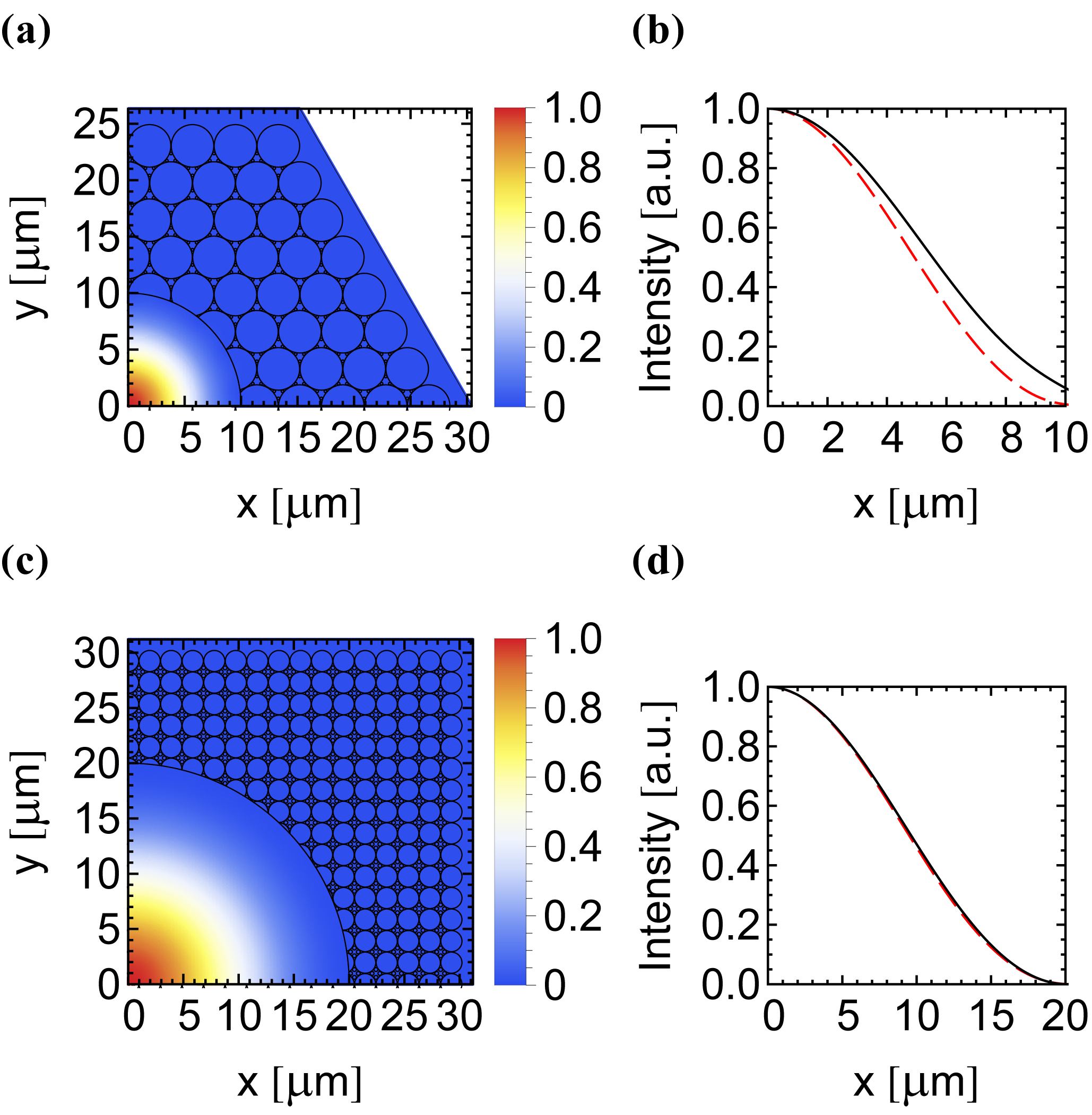}
\caption{\label{fig:5}(a,c) Spatial distributions of the third harmonic modes in the higher order bandgaps of both fiber designs. (b,d) The distribution of the modes in the fundamental (black) and higher order (red dashed) bandgaps along the ${x}$ axes in the hybrid hexagonal (b) and square (d) fiber cores.}
\end{figure}

In Fig.~\hyperref[fig:4]{\ref*{fig:4}}, we plot the confinement loss for both hybrid structures in the fundamental and higher-order bandgaps. These FEM calculations were performed with a commercial software (JCMwave), where the resolution of the FEM calculations is ${1~nm}$. Material dispersion was included with a three-term Sellmeier equation. We selected cladding periods of ${\Lambda_h=3.8~\mu m}$ and ${\Lambda_s=1.95~\mu m}$ to design both fibers for a ${\lambda=1.5~\mu m}$ fundamental wavelength source. With these parameters, the strut thicknesses and air hole radiuses for the hybrid hexagonal fiber are ${t_h^1=38~nm}$, ${t_h^2=19~nm}$, ${r_h^1=1.881~\mu m}$ and ${r_h^2\sim0.1159~\mu m}$, respectively. Similarly, the parameters of the hybrid square fiber are ${t_s^1=39~nm}$, ${t_s^2=29.25~nm}$, ${r_s^1=0.9555~\mu m}$ and ${r_s^2\sim0.1724~\mu m}$. The periodic background simulated in the hybrid hexagonal fiber has 7 hexagon rings of large air holes with a ${20~\mu m}$ core diameter~(Fig.~\hyperref[fig:5]{\ref*{fig:5}(a)}), and the hybrid square fiber has a periodic background array of 31x31 large air holes with a ${40~\mu m}$ core diameter~(Fig.~\hyperref[fig:5]{\ref*{fig:5}(c)}). The core wall thickness was neglected in these calculations to show the overall behavior of the confinement loss. The finiteness of the core wall supports more surface modes and reduces the transmission bandwidth of the fiber. Anti-crossing of the fundamental mode with surface modes causes the confinement loss to be discontinuous and increases the loss at these crossings. A state-of-the-art thin core wall
design~\cite{poletti_towards_2013} can move surface modes to the edges of the bandgaps and help to exploit the full low-loss spectral range of the \mbox{HC-PBG}. These thin core walls that are as a rule of thumb approximately half the size of the cladding struts, may also improve the spectral bandwidth in our designs; however, this must be confirmed numerically. \mbox{HC-PBG} core walls with strut thicknesses similar to those in both cladding designs proposed here have been fabricated successfully in the past~\cite{poletti_towards_2013}. Furthermore, a fiber with a~${\sim110~nm}$ air core diameter, which is almost half the size of the smallest air hole diameter in both cladding layouts, has been fabricated~\cite{wiederhecker_field_2007}. Yet, there has not been any effort to fabricate highly symmetric \mbox{HC-PBG} with such thin struts in the cladding; thus, the fabrication of these proposed \mbox{HC-PBGs} should be feasible, although not an easy or straightforward task to accomplish.

The spectral widths of the fundamental bandgap in the hybrid hexagonal and square fiber structures are ${\sim540~nm}$ and ${\sim240~nm}$, respectively. Both fibers have a higher-order bandgap with a spectral width of ${\sim30~nm}$. The central wavelengths of the higher-order bandgaps are slightly~${(<11~nm)}$ shifted from the designated ${0.5~\mu m}$ third harmonic wavelength. This shift can be compensated by shifting the pump source as well. The confinement loss is below ${1~dB/km}$ at the center of both the fundamental and higher-order bandgaps of the hybrid hexagonal fiber and also at the center of the higher-order bandgap of the hybrid square fiber. As a consequence, the total loss in these bandgaps is expected to be dominated by scattering losses~\cite{roberts_ultimate_2005}. The highest confinement loss is in the fundamental bandgap of the square structure, and the loss at the center of this specific bandgap is ${\sim366~dB/km}$.

Spatial distributions of the fundamental and third harmonic modes in the hybrid hexagonal and square fiber cores are illustrated in Fig.~\hyperref[fig:5]{\ref*{fig:5}}. The modal overlap appears to be exceptionally good for both fiber layouts. To determine the fraction of intensity in the hollow-core of each fiber design, we interpolated the intensity of the fundamental and higher-order modes to a 3rd degree polynomial along the ${x}$ axes. Only a small fraction of energy penetrates into the cladding. Both the fundamental and higher-order modes of the hybrid square fiber, and also the higher-order mode of the hybrid hexagonal fiber, have~${\sim99.9\%}$ of their intensity in the core. The fundamental mode of the hybrid hexagonal fiber has a similar fraction of intensity in the core, namely,~${\sim99.1\%}$.

In conclusion, we present two novel dual-bandgap \mbox{HC-PBGs} that can support a fundamental mode and its third harmonic simultaneously. The good modal overlap and the possibility to guide both the pump and the third harmonic with low losses, constitute two of the three building blocks required for efficient THG in gas-filled \mbox{HC-PBGs}; still missing is the phase-matching between the pump and the high harmonic, which may be achieved with careful dispersion engineering of the fiber combined with gas dispersion. We plan to pursue this goal in future investigations.

\nocite{*}

\bibliography{apssamp}

\end{document}